\documentclass[twocolumn,aps,showpacs]{revtex4}
\usepackage[latin9]{inputenc}
\usepackage{amsmath}
\usepackage{amssymb}
\usepackage{graphicx}

\makeatletter
\@ifundefined{textcolor}{}
{%
 \definecolor{BLACK}{gray}{0}
 \definecolor{WHITE}{gray}{1}
 \definecolor{RED}{rgb}{1,0,0}
 \definecolor{GREEN}{rgb}{0,1,0}
 \definecolor{BLUE}{rgb}{0,0,1}
 \definecolor{CYAN}{cmyk}{1,0,0,0}
 \definecolor{MAGENTA}{cmyk}{0,1,0,0}
 \definecolor{YELLOW}{cmyk}{0,0,1,0}
}

\usepackage{color}

\usepackage{braket}

\def\NOT(#1,#2){\OneQubitGate(#1,#2){$X$}}

\makeatother

\begin{document}

\title{Bloch-Siegert Shift in a Hybrid Quantum Register: Quantification
and Compensation}

\author{Jingfu Zhang$^{1}$, Sagnik Saha$^{1,2}$ and Dieter Suter$^{1}$\\
 $^{1}$ Fakult$\ddot{a}$t Physik, Technische Universit$\ddot{a}$t Dortmund,\\
 D-44221 Dortmund, Germany\\
 $^{2}$ current address : Indian Institute of Science Education and
Research Kolkata, \\
 Mohanpur, India-741246 }

\date{\today}
\begin{abstract}
Quantum registers that combine the attractive properties of different
types of qubits are useful for many different applications. They also
pose a number of challenges, often associated with the large differences
in coupling strengths between the different types of qubits. One example
is the non-resonant effect that alternating electromagnetic fields
have on the transitions of qubits that are not targeted by the specific
gate operation. The example being studied here is known as Bloch-Siegert
shift. Unless these shifts are accounted for and, if possible, compensated,
they can completely destroy the information contained in the quantum
register. Here we study this effect quantitatively in the important
example of the nitrogen vacancy (NV) center in diamond and demonstrate how
it can be eliminated. 
\end{abstract}

\maketitle

\section{Introduction}

Storage and processing of information in quantum mechanical systems,
known as quantum information processing \cite{nielsen,Stolze:2008xy},
has an enormous potential for many applications where classical systems
can not provide sufficient computational power. The realization of
this potential relies, amongst others, on the capability to selectively
apply gate operations to specific quantum bits (qubits), without perturbing
the other qubits present in the system. In most systems that are currently
being studied for this type of applications, the selection of the
qubits is performed in frequency space: An alternating electric or
magnetic field whose frequency is tuned to the transition frequency
of the targeted qubit drives the target qubit in a way that can be
well described by a rotation on the Bloch sphere. This type of resonant
excitation works well for many different systems, such as electronic
and nuclear spins, trapped atomic ions or neutral atoms, but also
engineered systems like superconducting circuits \cite{ReviewQC10}.

The required selectivity of this frequency-domain addressing scheme
is typically assumed to work well if the Rabi frequency of the driving
field is small compared to the frequency difference between the qubits.
As an example, if two qubits have transition frequencies $\Omega_{A}$
and $\Omega_{B}$, the Rabi frequency $\omega_{1}$ should fulfill
the condition $|\omega_{1}|\ll|\Omega_{A}-\Omega_{B}|$ - a condition
that can often be fulfilled.

The situation becomes more complicated in systems where the coupling
strength between the qubits and the resonant field differs strongly
between the different qubits. A particularly important type of such
hybrid quantum registers are systems that consist of electronic and
nuclear spins. The coupling strength between a spin and a magnetic
field is quantified by the gyromagnetic ratio $\gamma$: $\omega_{1}=\gamma B_{1}$,
where $B_{1}$ is the amplitude of the AC magnetic field. If an alternating
magnetic field is used to drive the nuclear spin, the nuclear spin
Rabi frequency is $\omega_{1n}=\gamma_{n}B_{1}$. The same field interacts
also with the electron spin, and the corresponding Rabi frequency
is $\omega_{1e}=|\gamma_{e}|B_{1}$, where $\gamma_{e}=-28$ GHz/T is
the electronic gyromagnetic ratio, which is some three orders of magnitude
larger than $\gamma_{n}$. In most cases, the condition $|\omega_{1e}|\ll|\Omega_{e}-\Omega_{n}|\approx|\Omega_{e}|$
is still well fulfilled and accordingly the resonant excitation of
the electron spin is negligibly small. However, since the nuclear
spin responds only slowly to the driving field, the gate operation
is also several orders of magnitude longer than the electron spin
operations. For such long pulse durations, also non-resonant effects
can become relevant, in particular the Bloch-Siegert (BS) shift \cite{PhysRev.57.522,PhysRev.138.B979,chemphyslett168297}:
This effect can be described by an effective Hamiltonian for the transition
where it is observed: 
\begin{equation}
\frac{1}{2\pi}\mathcal{H}_{BS}=\omega_{BS}\frac{\sigma_{z}}{2},\label{BSHam}
\end{equation}
where $\sigma_{z}$ denotes the $z$-component of the Pauli matrix.
The frequency shift 
\begin{equation}
\omega_{BS}=\frac{\omega_{1}^{2}}{2\omega_{0}}\label{BSthrtext}
\end{equation}
is given by the square of the Rabi frequency $\omega_{1}$, divided
by twice the transition frequency $\omega_{0}$. It can be calculated
by the Floquet formalism \cite{PhysRev.138.B979}, with a dressed
atom model \cite{0022-3700-6-8-007} or a unitary approach \cite{PhysRevA.91.053834}.
Here $\omega_{0}$ denotes the transition frequency of the electron
spin. The effect has been observed, e.g. in pulsed ENDOR \cite{3351},
circuit quantum electrodynamics systems \cite{PhysRevLett.105.237001,PhysRevB.93.214501,PhysRevB.96.020501},
and Rydberg atoms \cite{PhysRevA.69.031401}.

Clearly, this effect becomes large when strong pulses are used, such
as in the ultra-strong coupling regime \cite{RabiAnn13,Nature458178,NatPhys5105,NatPhys6772,PhysRevA.95.053804,NatPhys1339,NatPhys1344,NatComm8779,PhysRevLett.105.257003}.
In the case of high-fidelity quantum control, the effects are smaller
but no less relevant, since they may significantly degrade the fidelity
of the operations \cite{BSJMRChuang,PhysRevA.78.012328}. Here, we
investigate the BS effects that RF pulses have on the electron spin
of a nitrogen vacancy (NV) center in diamond.

\section{System and Experimental Protocol}

We demonstrate the issue for the example of a system consisting of
the electron and $^{14}$N nuclear spins in a single 
NV center , subjected to a magnetic field $B$ oriented along its
symmetry axis. The Hamiltonian for this system can be written as ~\cite{PhysRevB.89.205202,Suter201750}
\begin{equation}
\frac{1}{2\pi}\mathcal{H}=DS_{z}^{2}-\gamma_{e}BS_{z}+PI_{z}^{2}-\gamma_{n}BI_{z}+AS_{z}I_{z}.\label{Hamsim}
\end{equation}
Here $S_{z}$ and $I_{z}$ are the $z$-components of the spin-1 operators
for electronic and nuclear spins, respectively. The zero-field splitting
is $D=2.87$ GHz, the nuclear quadrupolar splitting $P=-4.95$ MHz,
the hyperfine coupling $A=-2.16$ MHz ~\cite{PhysRevB.89.205202,PhysRevB.47.8816,Yavkin16}
and the nuclear gyromagnetic ratio $\gamma_{n}=3.1$ MHz/T. In the
experiments, the static field strength is about $15$ mT, which results
in a separation of the two electron spin resonance (ESR) transitions
by about $840$ MHz.

\begin{figure}
\centering{}\includegraphics[width=1\columnwidth]{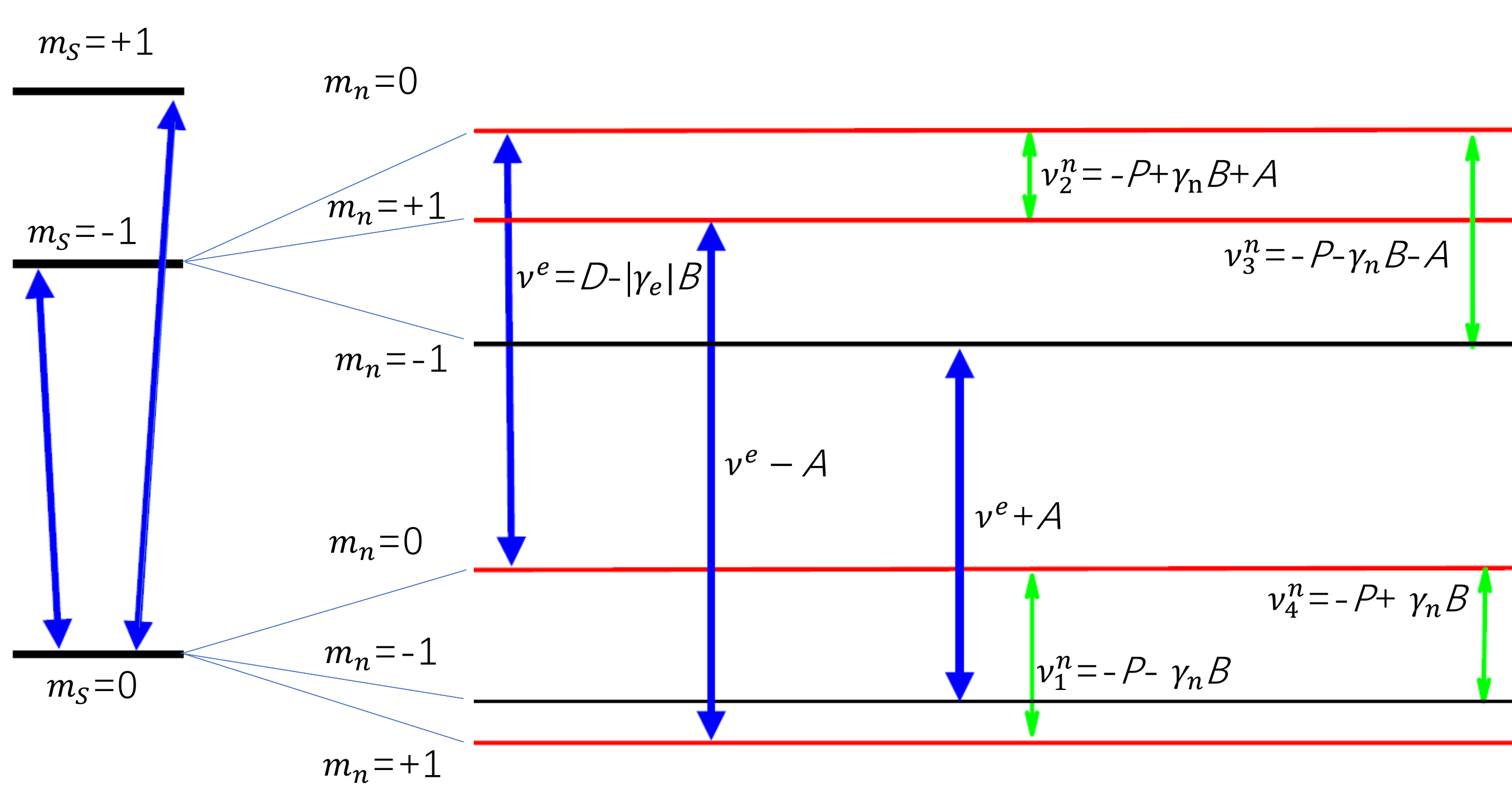}
\caption{(color online). Energy diagram of the NV center coupled with $^{14}$N spin. The four
basis states marked by red horizontal lines form a 2-qubit system.
The vertical arrows indicate the allowed ESR and NMR transitions.
The transition frequencies are indicated next to the arrows. \label{figenergy}}
\end{figure}

Figure \ref{figenergy} shows the corresponding energy level scheme,
focusing on the case where we excite the $m_{S}=0\leftrightarrow-1$
transitions of the electron spin. The ESR transition frequency was
$\nu_{1}^{e}=2438.739$ MHz, the NMR transition frequencies were $\nu_{1}^{n}=4.990$,
$\nu_{2}^{n}=2.828$, $\nu_{3}^{n}=7.066$, and $\nu_{4}^{n}=4.898$
MHz, and the nuclear spin Rabi frequencies were 10.7, 6.0, 6.3, and
10.4 kHz at an RF power of $p_{0}$$\approx$ 80 mW. As a minimal
system for the purpose of this demonstration, we focus on the four
levels marked by thick red lines, which form a 2-qubit system.

\begin{figure}
\centering{}\includegraphics[width=0.8\columnwidth]{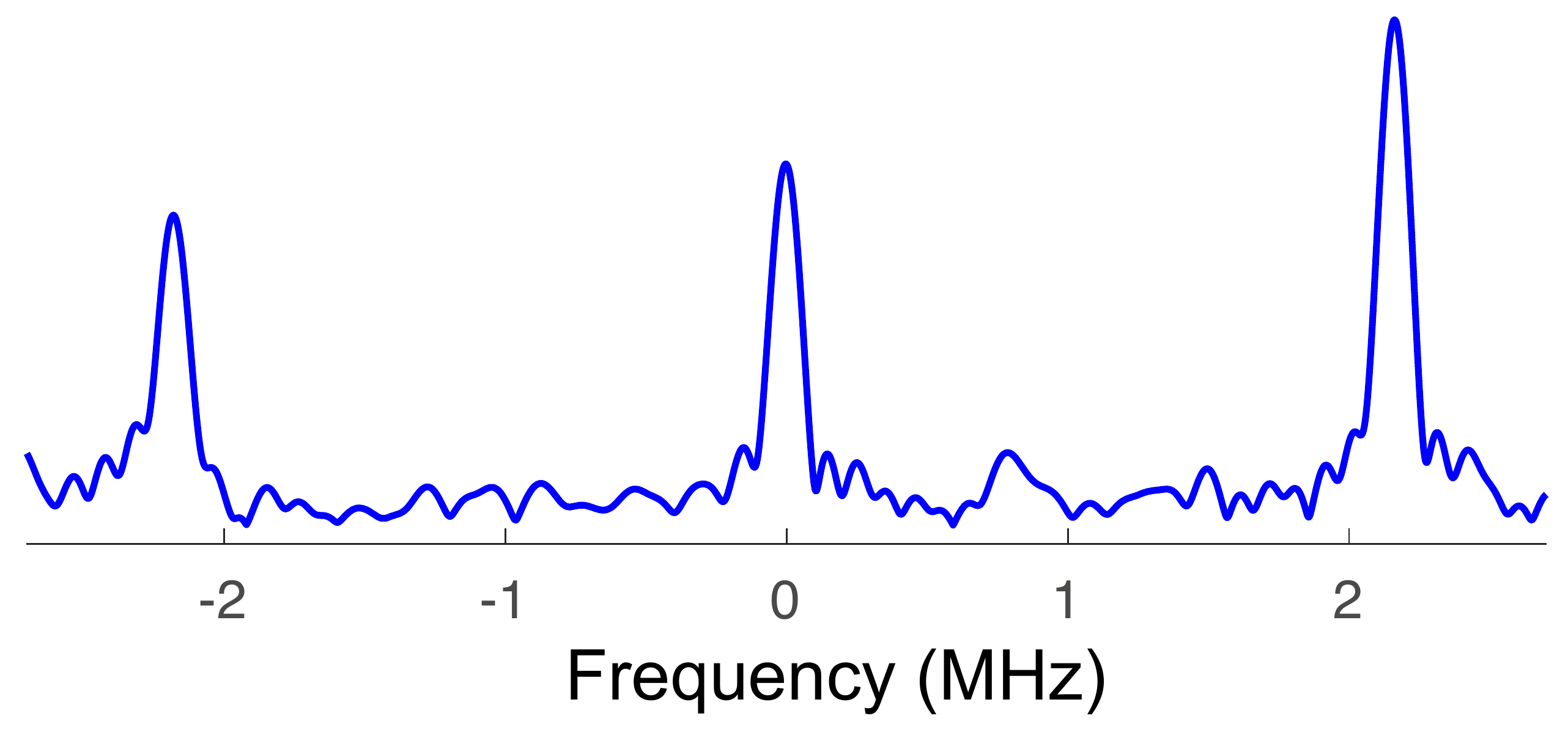} 
\caption{(color online). Spectrum of the ESR transitions between the states with $m_{s}=0$
and $-1$, obtained as Fourier-transform of the time-domain signals.
The origin of the frequency axes is set to $D-|\gamma_{e}|B$, measured
as 2438.739 MHz. \label{figFFT}}
\end{figure}

Figure \ref{figFFT} shows the spectra of the ESR transitions between
the states with $m_{s}=0$ and $-1$, obtained in a Ramsey-type free-induction
decay (FID) experiment, using resonant microwave (MW) pulses with Rabi
frequencies of about 10 MHz for excitation and detection.

\begin{figure}
\centering{}\includegraphics[width=1\columnwidth]{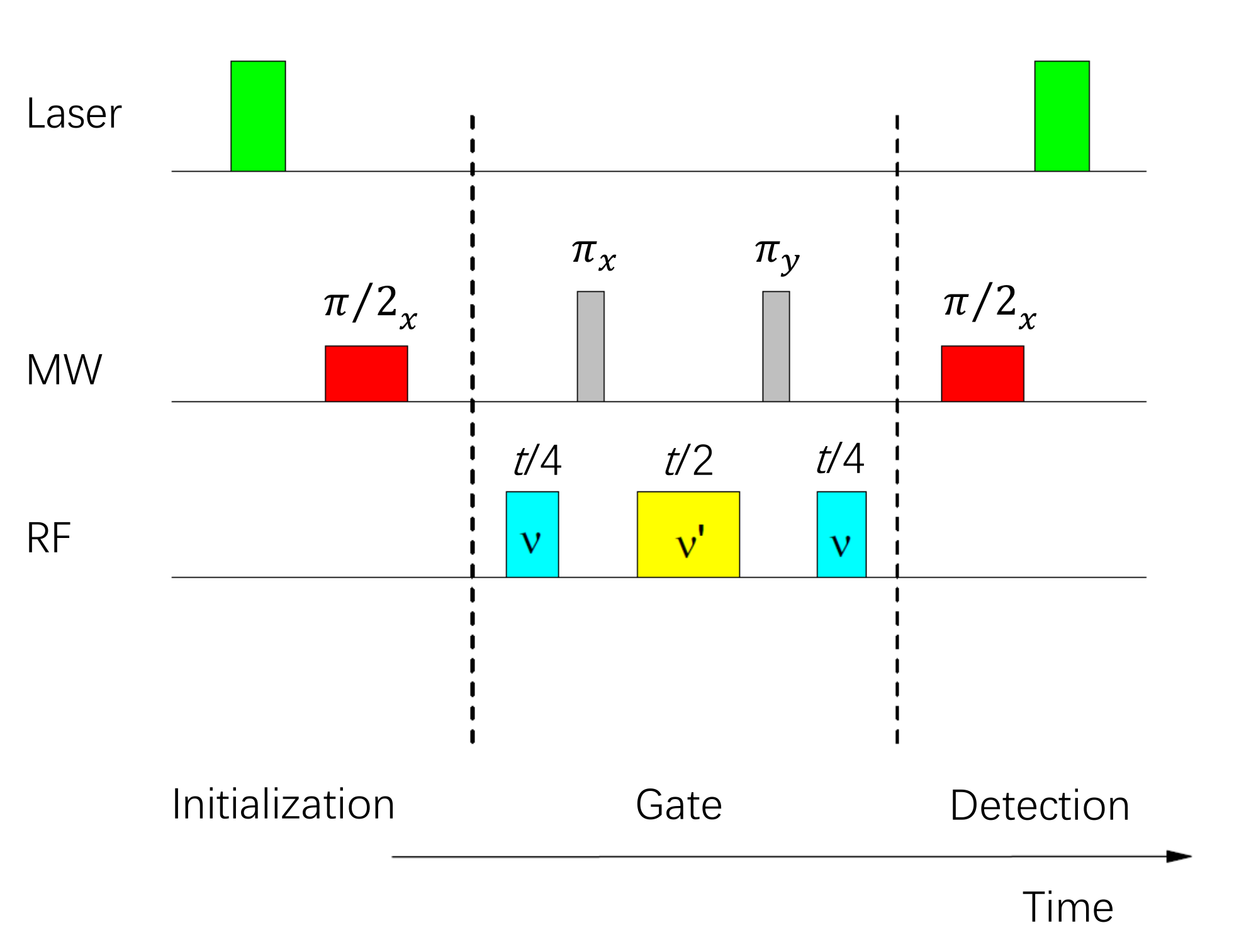}
\caption{(color online). Pulse sequence. The laser pulses ($\lambda=$ $532$ nm) are used
for initialization and detection of the electron spin. The MW pulses are resonant with the transition $|0\rangle_{e}|0\rangle_{n}\leftrightarrow|-1\rangle_{e}|0\rangle_{n}$.
In  the initialization and detection steps, they are either transition-selective
or hard pulses with Rabi frequency of about $0.30$ MHz or 12 MHz,
respectively. The flip angles are $\pi/2$, with the phase indicated
by the index. During the gate, the MW pulses are hard pulses with
Rabi frequency about $12$ MHz. The frequencies of the RF pulses are indicated in the rectangles. \label{figBSpulse1}}
\end{figure}

Figure \ref{figBSpulse1} shows the pulse sequence used to demonstrate
the effect and its compensation. During the initialization, we use
a laser pulse to bring the electron spin to the $m_{s}=0$ state,
and then a MW pulse with a $\pi/2$ flip angle to generate
coherence between the $m_{s}=0$ and $-1$ states. This coherence
allows us to observe the effect of the BS shift, which is generated
by the radio frequency (RF) pulses during the period marked as ``gate''
in figure \ref{figBSpulse1}. These RF pulses are required to drive
gate operations on the $^{14}$N nuclear spin. The color (blue, yellow)
of the pulses and the frequency labels ($\nu$, $\nu'$) of the pulses
indicate, with which transition they are resonant. During the gaps
in the RF pulses, we apply dynamical decoupling (DD) pulses to the
electron spin, to refocus unwanted dephasing of the electron spin
coherence. During the detection period, the MW pulse transforms the
coherence to spin population and the laser pulse reads out the population
of the $m_{S}=0$ state $P_{|0\rangle}$.

\section{Measurement of the Frequency Shift}

The applied RF field is oriented at an arbitrary direction with respect
to the coordinate system of the NV center and has therefore components
parallel as well as perpendicular to the NV-axis. The parallel component,
which corresponds to the $z$-component in the conventional choice
of coordinate system, leads to oscillations with the frequency of
the RF field \cite{nature17404}. Here, we focus on the transverse
components, which generate the BS-shift. To quantitate the observed
effect and verify the interpretation as a BS shift, we measured the
dependence of the shift on the the RF-amplitude. For this purpose,
we applied RF pulses only before the first and after the second DD
pulse. In the initialization and detection steps, we used transition
selective MW pulses with the carrier frequency set to the transition
$|0\rangle_{e}|0\rangle_{n}\leftrightarrow|-1\rangle_{e}|0\rangle_{n}$.

The $\pi/2$ MW pulse in the initialization step generates a superposition
of the electron spin 
\begin{equation}
|s_{0}\rangle=(|0\rangle-i|1\rangle)/\sqrt{2}.\label{state0}
\end{equation}
The $z$ -rotation by the Hamiltonian (\ref{BSHam}) in the gate step
\begin{equation}
U_{BS}(t)=e^{-i2\pi\omega_{BS}t\sigma_{z}/4},\label{Ueff}
\end{equation}
turns $|s_{0}\rangle$ into 
\begin{equation}
|s(t)\rangle=(|0\rangle e^{-i2\pi\omega_{BS}t/4}-i|1\rangle e^{i2\pi\omega_{BS}t/4})/\sqrt{2}.\label{statetime}
\end{equation}
During the detection step, the second $\pi/2$ MW pulse transforms
$|s(t)\rangle$ to the final state 
\begin{equation}
|s_{f}(t)\rangle=|0\rangle\cos(2\pi\omega_{BS}\frac{t}{4})+|1\rangle\sin(2\pi\omega_{BS}\frac{t}{4}),\label{statet}
\end{equation}
where the population $P_{|0\rangle}$ of $|0\rangle$ encodes the
BS shift.

\begin{figure}
\centering{}\includegraphics[width=1\columnwidth]{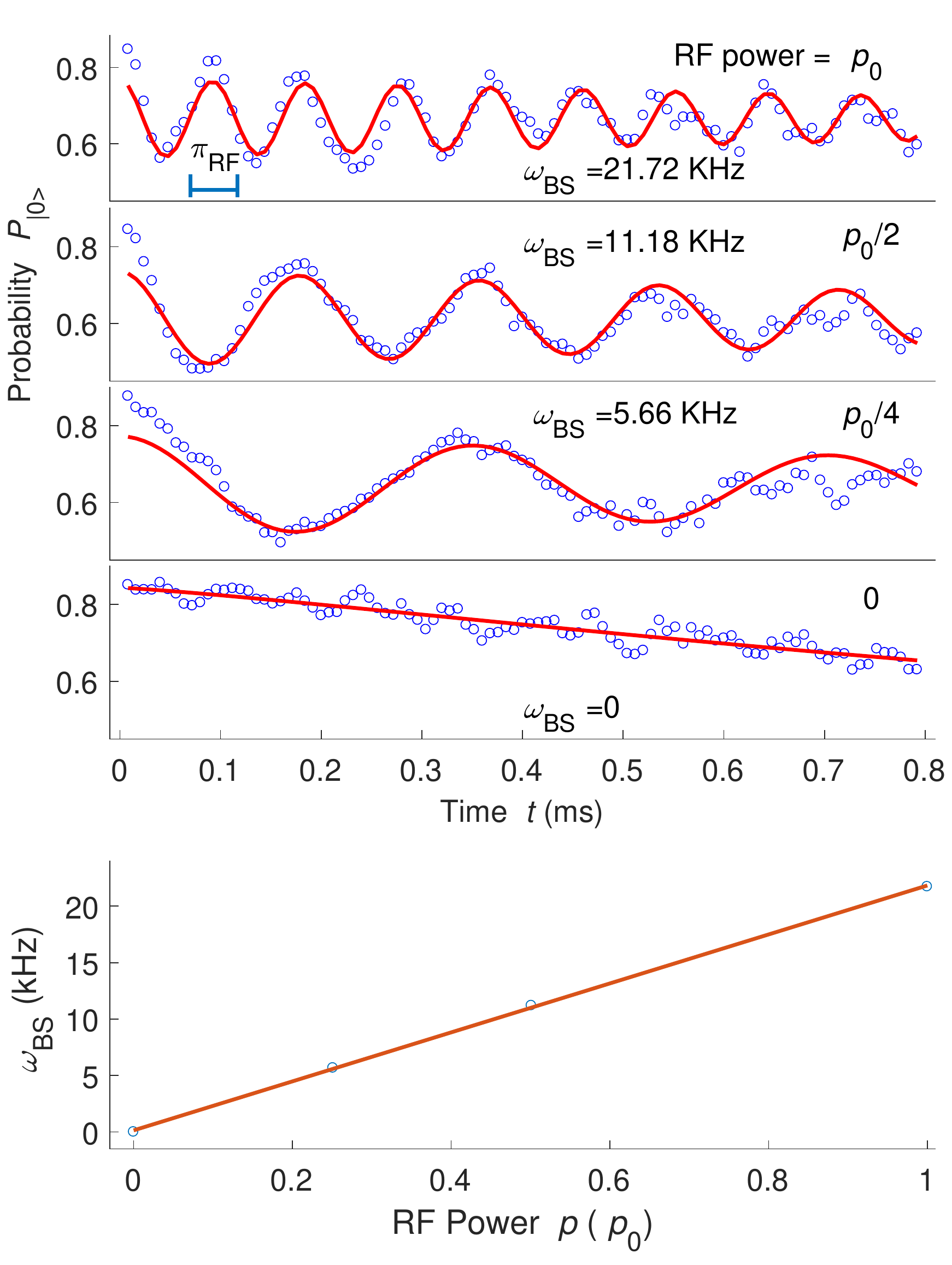}
\caption{(color online). Experimental results for the dependence of the BS shift on the amplitude
$B_{1}$ of the RF pulses. The RF frequency was 6 MHz and the RF power
was $p_{0}$, $p_{0}/2$, and $p_{0}/4$ and 0. The horizontal scale
bar in the panel for RF power $p_{0}$ indicates the duration of the
$\pi$ pulse for the $^{14}$N spin when the RF frequency is resonant
with the transition at 4.990 MHz. The experimental data are shown
as empty circles. The fit to the data, shown as the solid curves,
gave the frequencies $\omega_{BS}=21.72$, 11.18 and 5.66 kHz. The
bottom figure shows the dependence of $\omega_{BS}$ on the RF power,
with the measured data as empty circles, and a linear fit to the data.
\label{figBSres1}}
\end{figure}

Figure \ref{figBSres1} shows the experimental results. for RF powers
from zero to 80 mW. In the absence of an RF field, the electron spin
coherence decays exponentially and we could fit the experimental data
with the function 
\begin{equation}
P_{|0\rangle}=a_{1}+b_{1}e^{-(t/T_{2})^{k}},\label{fitDD}
\end{equation}
and obtained $T_{2}$ $=$ 1.3 ms and $k=1.2$.

To test the effect of the RF field, we first set the frequency far
from the nuclear spin transitions and applied RF power during the
first and third period of the gate (blue pulses in Fig. \ref{figBSpulse1}).
The top three traces of Figure \ref{figBSres1} show the resulting
signals from the electron spin, measured with an RF frequency of $\nu=6$
MHz and RF powers of $p_{0}=\mathrm{80\,mW}$, $p_{0}/2$, and $p_{0}/4$.
The data can be fitted as 
\begin{equation}
P_{|0\rangle}=a_{2}+b_{2}\cos(2\pi\frac{\omega_{BS}}{2}t)e^{-(t/T_{2})}.\label{fitBS1}
\end{equation}
The resulting fit parameters are $\omega_{BS}=21.72$, $11.18$ and
$5.66$ kHz for decreasing power. The bottom trace of figure \ref{figBSres1}
shows the dependence of the measured frequency on the RF power. According
to Eq. (\ref{BSthrtext}), the dependence should be linear (quadratic
in $B_{1}$), which is well borne out by the experimental data.

\begin{figure}
\centering{}\includegraphics[width=1\columnwidth]{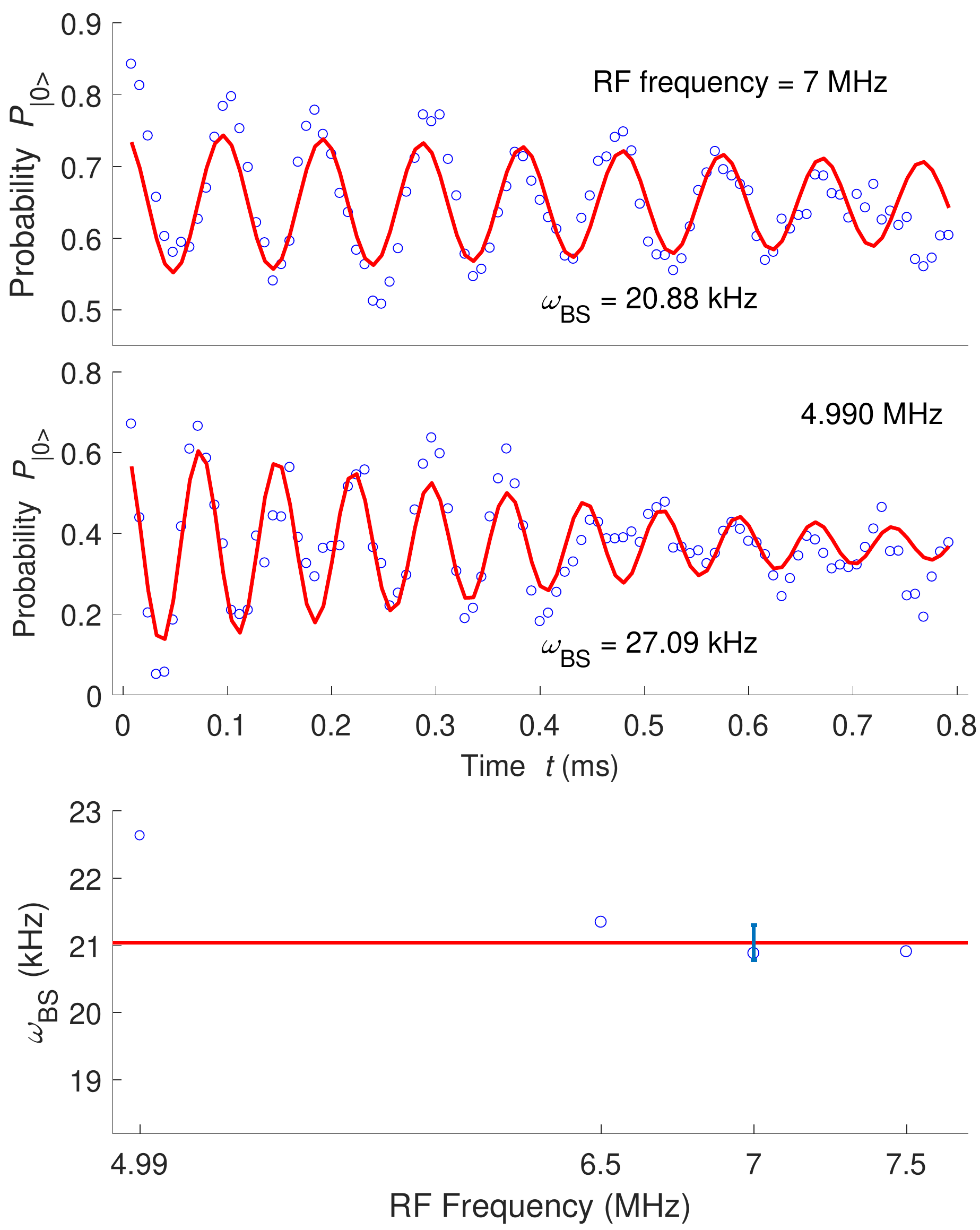}
\caption{(color online). Measurement of the BS shift as a function on the frequency of the
RF pulses, obtained with the first and third RF pulses at power $p_{0}$. The experimental data are shown as circles and
the RF frequencies are indicated in the panels. By fitting the data,
shown as the solid curves, we obtained the BS shift indicated in the
panel. In the bottom figure, we show the measured BS shift by the
empty circles for RF frequencies 6.5, 7, and 7.5 MHz. The solid line
shows the average, and the error bar shows the standard deviation.
Since the RF power was not constant as a function of frequency, we
normalized the shift observed at 4.99 MHz to the same RF power as
the others. \label{figBSres2}}
\end{figure}

In the second set of experiments, we fixed the power of the RF pulses
to $p_{0}$ and varied the frequency from 6.5 MHz to 7 and 7.5 MHz.
By fitting the data using function (\ref{fitBS1}), with different
constant $a_{2}$ and $b_{2}$, we obtained the oscillation frequency
as $\omega_{BS}=21.34$, $20.88$ and $20.90$ kHz, as illustrated
in Figure \ref{figBSres2}. Within the experimental uncertainty, the
oscillation frequency does not depend on the RF frequency, as shown
in the bottom part of Figure \ref{figBSres2}. This result is also
consistent with the prediction of Eq. (\ref{BSthrtext}).

For the case that the RF pulses are resonant with the NMR transitions,
they affect the nuclear as well as the electronic spin. In order to
eliminate this effect and observe only the BS shift, we use hard pulses
in the initialization and detection steps. In this case, we can neglect
the small polarization of the nuclear spin, shown as the small difference
in the peak amplitudes of Figure \ref{figFFT}. We applied RF pulses
with frequency $\nu=\nu_{1}^{n}$ = 4.990 MHz. The results are shown
in Figure \ref{figBSres2}. By fitting the data using the function
in Eq. (\ref{fitBS1}), we obtained $\omega_{BS}=27.09$ kHz. The
deviation from the results in the off-resonant case can be mainly
attributed to the difference of the powers. We therefore normalized
the measured $\omega_{BS}$ to the applied power;  the result is shown
in the bottom part of Figure \ref{figBSres2}. It is
close to the average value measured in the off-resonant case. 

\section{Compensation}

Since the BS shift corresponds to a quasi-static shift of the resonance
frequency, it can be eliminated by applying refocusing pulses, provided
it is constant for the whole period. For this demonstration, we use
a very simple DD sequence consisting of one $\pi_{x}$ and one $\pi_{y}$
pulse, as shown in Fig. \ref{figBSpulse1}. The total phase generated
by the RF pulses, which are applied only between the DD pulses, is
then $\phi_{BS}=\phi_{1}-\phi_{2}+\phi_{3}$, where the three terms
denote the phase acquired before the first, between the two and after
the second DD pulse. If the RF power is constant and the relative
pulse durations are 1:2:1, the total phase vanishes, $\phi_{BS}=0$.

\begin{figure}
\centering{}\includegraphics[width=1\columnwidth]{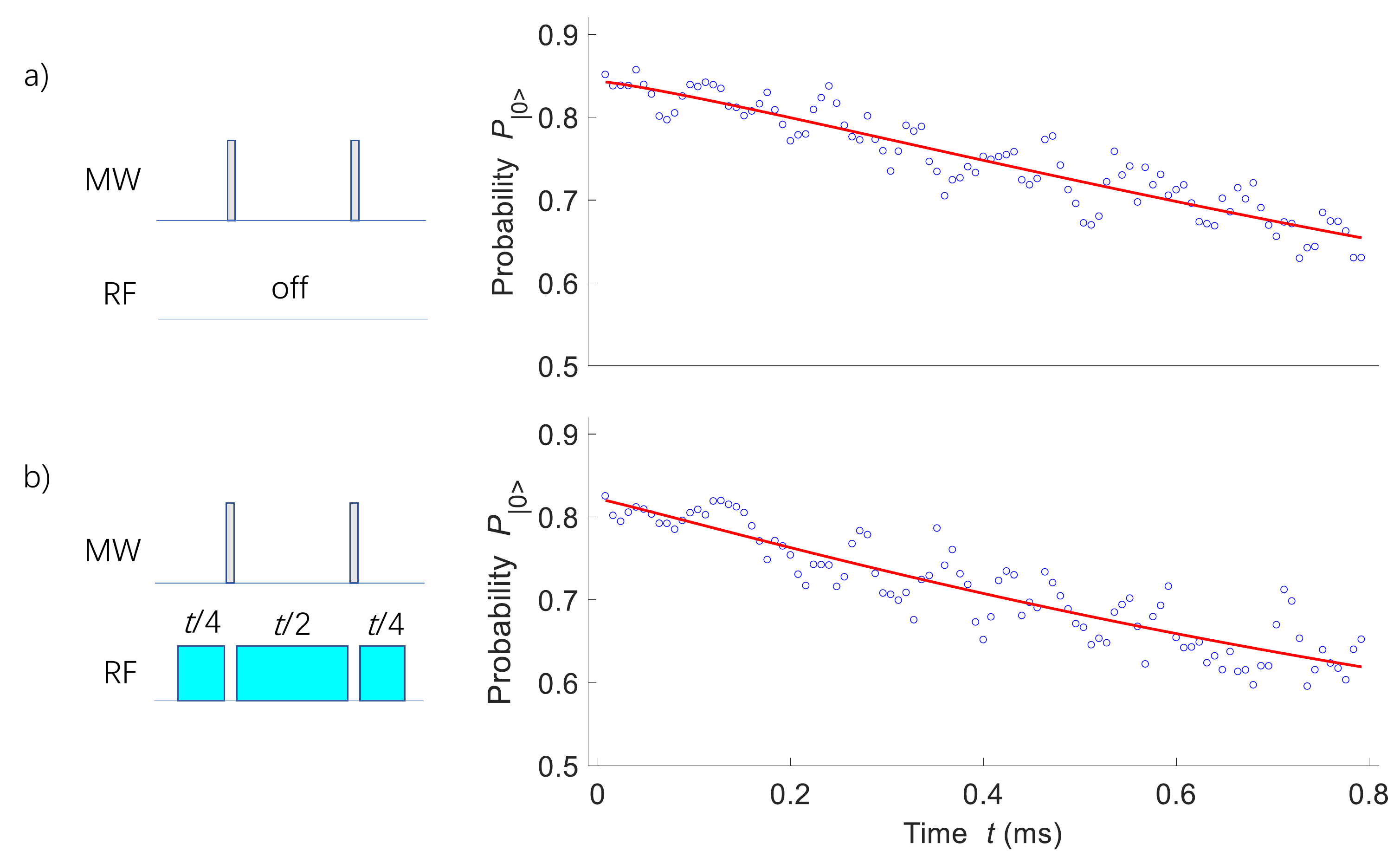}
\caption{(color online). Cancellation of the BS shift by DD pulses, indicated by the grey narrow
rectangles. The experimental data are indicated by the empty circles,
and the fits with Eq. (\ref{fitDD}) by solid curves. (a) Without RF: no shift, decay of the
coherence with $T_{2}=1.3$ ms and $k=1.2$. (b) All the RF pulses
in the sequence are switched on and have the same frequency (6 MHz)
and power $p_{0}$. The experimental data can be fitted
with $T_{2}=1.2$ ms and $k=1.1$.\label{figBScancle}}
\end{figure}

In Figure \ref{figBScancle}, we illustrate the experimental results
in the case of off-resonant RF pulses with power $p_{0}$ and frequency
$\nu^{'}=\nu=6$ MHz, and compare it to the case without RF. In both
cases, we observe no oscillations, indicating that the phases generated
by the different pulses cancel each other completely. We can use function
(\ref{fitDD}) to fit the non-oscillatory data, and obtained the fitted
$T_{2}=1.2$ and 1.3 ms for data with and without RF pulses. The similarity
between the results with and without RF pulses shows the good cancellation
of the BS shift, consistent with the prediction in previous work \cite{nature17404}.

\section{discussion}

Since the BS shift is sensitive to the RF power but independent of
the RF frequency, it might be a valuable resource to measure or calibrate
the RF power using the electron spin as a probe. One potential application
could be for estimating the perpendicular component $A_{\perp}$ of
the hyperfine coupling from $^{14}$N in the NV center system. In principle,
one can extract $A_{\perp}$ from the measured Rabi frequencies of
$^{14}$N, using the different enhancement for different states of
the $^{14}$N \cite{PhysRevB.92.020101}. However this strategy requires
that the response of the electronics to a change of the RF frequency
is known precisely. Although this response can be partially corrected
by measuring the RF power, the deviation is still not small enough,
shown as the data in the bottom plot in Figure \ref{figBSres2}.

In order to normalize the RF power for measuring Rabi frequencies
of $^{14}$N, we can set an offset, e.g., 0.5 MHz from the concerned
NMR transition, noting that the RF power in such a range can be treated
as a constant, shown as the experimental data in
the bottom plot in Figure \ref{figBSres2}. Since the maximal Rabi
frequencies are about 10 kHz in our experiment, the offset of 0.5 MHz
is large enough to avoid the resonance effect from the NMR transition.
Therefore we can normalize the RF power by measuring the BS effect
using the protocol proposed in the current paper. Then we can extract
$A_{\perp}$ from the corrected Rabi frequencies. 

\section{Conclusion}

In conclusion, we have demonstrated that AC magnetic fields that are
used to drive the nuclear spins in a hybrid quantum register can have
a large effect on the electronic spin although their frequency is
very far from their resonance frequencies. The main effect is a phase
shift, which can easily exceed $2\pi$ and therefore completely scramble
the quantum information stored in the electron spin qubit, unless
measures are taken to eliminate its effect. The purpose of this work
was to quantitatively measure the effect and test schemes for compensation.
We have analyzed the dependence of the BS shift on the power and frequency
of the RF pulses and demonstrated that spin echoes can refocus the
BS effect with excellent precision.

\section{acknowledgement}
This work was supported by the DFG through grant 192/34-1. S. Saha thanks the support by DAAD-WISE 2017 for financing his travel and stay in Germany to carry out the research project. We also thank Dr. Swathi Hegde for useful discussions. 
 \bibliographystyle{apsrev}

\begin{thebibliography}{}
\bibitem {nielsen} M. A. Nielsen and I. L. Chuang, {\it Quantum Computation
and Quantum Information} (Cambridge University Press,
Cambridge, 2000).

\bibitem {Stolze:2008xy}  J. Stolze and D. Suter, {\it Quantum Computing: A Short
Course from Theory to Experiment} (Wiley-VCH, Berlin, 2nd edition, 
2008).

\bibitem {ReviewQC10} T. D. Ladd, F. Jelezko, R. La
amme, Y. Nakamura,
C. Monroe, and J. L. O'Brien, Nature {\bf 464}, 45 (2010).

\bibitem {PhysRev.57.522} F. Bloch and A. Siegert, Phys. Rev. {\bf 57}, 522 (1940).

\bibitem {PhysRev.138.B979} J. H. Shirley, Phys. Rev. {\bf 138}, B979 (1965).

\bibitem {chemphyslett168297}  L. Emsley and G. Bodenhausen, Chem. Phys. Lett. {\bf 168},
297 (1990). 

\bibitem {0022-3700-6-8-007} C. Cohen-Tannoudji, J. Dupont-Roc, and C. Fabre, Journal
of Physics B: Atomic and Molecular Physics {\bf 6}, L214
(1973). 

\bibitem {PhysRevA.91.053834} Y. Yan, Z. Lv, and H. Zheng, Phys. Rev. A {\bf 91},
053834 (2015).

\bibitem {3351} M. Mehring, P. Hoefer, and A. Grupp, Phys. Rev. A {\bf 33},
3523 (1986).

\bibitem {PhysRevLett.105.237001} P. Forn-Diaz, J. Lisenfeld, D. Marcos, J. J. Garcia-Ripoll,
E. Solano, C. J. P. M. Harmans, and J. E. Mooij, Phys. Rev. Lett. {\bf 105}, 237001 (2010).

\bibitem {PhysRevB.93.214501}  A. Baust, E. Hoffmann, M. Haeberlein, M. J.
Schwarz, P. Eder, J. Goetz, F. Wulschner, E. Xie,
L. Zhong, F. Quijandria, et al., Phys. Rev. B {\bf 93},
214501 (2016).

\bibitem {PhysRevB.96.020501} I. Pietikainen, S. Danilin, K. S. Kumar, A. Vepsalainen,
D. S. Golubev, J. Tuorila, and G. S. Paraoanu, Phys.
Rev. B {\bf 96}, 020501 (2017).

\bibitem {PhysRevA.69.031401} D. Fregenal, E. Horsdal-Pedersen, L. B. Madsen,
M. F{\o}rre, J. P. Hansen, and V. N. Ostrovsky, Phys. Rev.
A {\bf 69}, 031401 (2004).

\bibitem {RabiAnn13} A. Moroz, Annals of Physics {\bf 340}, 252 (2014).

\bibitem {Nature458178} G. Gunter, A. A. Anappara, J. Hees, A. Sell, G. Biasiol,
L. Sorba, S. D. Liberato, C. Ciuti, A. Tredicucci,
A. Leitenstorfer, et al., Nature {\bf 458}, 178 (2009).

\bibitem {NatPhys5105} L. S. Bishop, J. M. Chow, J. Koch, A. A. Houck,
M. H. Devoret, E. Thuneberg, S. M. Girvin, and R. J.
Schoelkopf, Nature Physics {\bf 5}, 105 (2009).

\bibitem {NatPhys6772} T. Niemczyk, F. Deppe, H. Huebl, E. P. Menzel,
F. Hocke, M. J. Schwarz, J. J. Garcia-Ripoll, D. Zueco,
T. Hummer, E. Solano, et al., Nature Physics {\bf 6}, 772
(2010).

\bibitem {PhysRevA.95.053804} K. R. K. Rao and D. Suter, Phys. Rev. A {\bf 95},
053804 (2017).

\bibitem {NatPhys1339} P. Forn-Diaz, J. J. Garcia-Ripoll, B. Peropadre, J.-L. Orgiazzi,
M. A. Yurtalan, R. Belyansky, C. M. Wilson, and
A. Lupascu, Nature Physics {\bf 13}, 39 (2017).

\bibitem {NatPhys1344} F. Yoshihara, T. Fuse, S. Ashhab, K. Kakuyanagi,
S. Saito, and K. Semba, Nature Physics {\bf 13}, 44 (2017).

\bibitem {NatComm8779}, J. Braumuller, M. Marthaler, A. Schneider, A. Stehli,
H. Rotzinger, W. Martin, and A. V. Ustinov, Nature
Communications {\bf 8}, 779 (2017).

\bibitem {PhysRevLett.105.257003} J. Tuorila, M. Silveri, M. Sillanpaa, E. Thuneberg,
Y. Makhlin, and P. Hakonen, Phys. Rev. Lett. {\bf 105}, 257003 (2010).

\bibitem {BSJMRChuang} M. Steffen, L. M. K. Vandersypen, and I. L. Chuang,
Journal of Magnetic Resonance {\bf 146}, 369 (2000).

\bibitem {PhysRevA.78.012328} C. A. Ryan, C. Negrevergne, M. Laforest, E. Knill,
and R. Laflamme, Phys. Rev. A {\bf 78}, 012328 (2008).

\bibitem {PhysRevB.89.205202} C. S. Shin, M. C. Butler, H.-J. Wang, C. E. Avalos, S. J.
Seltzer, R.-B. Liu, A. Pines, and V. S. Bajaj, Phys. Rev.
B {\bf 89}, 205202 (2014).

\bibitem {Suter201750} D. Suter and F. Jelezko, Progress in Nuclear Magnetic
Resonance Spectroscopy {\bf 98-99}, 50 (2017)

\bibitem {PhysRevB.47.8816} X.-F. He, N. B. Manson, and P. T. H. Fisk, Phys. Rev.
B {\bf 47}, 8816 (1993).

\bibitem {Yavkin16} B. Yavkin, G. Mamin, and S. Orlinskii, J. Magn. Reson.
{\bf 262}, 15 (2016).

\bibitem {nature17404} M. Hirose and P. Cappellaro, Nature {\bf 532}, 77 (2016).

\bibitem {PhysRevB.92.020101} M. Chen, M. Hirose, and P. Cappellaro, Phys. Rev. B
{\bf 92}, 020101 (2015)

\end{thebibliography}

\end{document}